\documentclass[10pt]{emulateapj}

\usepackage{times}

\usepackage{color}

\newcommand{\htwo}{H$_2$}

\newcommand{\msun}{M$_\odot$}
\newcommand{\kms}{km~s$^{-1}$\relax}

\newcommand{\degree}{$^{\circ}$}
\newcommand{\iizw}{II~Zw~40}
\newcommand{\gtd}{{\small GTD}}
\newcommand{\rthreeone}{$R_{31}$}
\newcommand{\jpbkms}{Jy~beam$^{-1}$~km~s$^{-1}$}
\newcommand{\jpb}{Jy~beam$^{-1}$}

\definecolor{mygreen}{rgb}{.05, .6, .01}
\definecolor{roxa}{rgb}{.5, .02, .6}

\begin{document}

\title{ALMA Reveals Potential Localized Dust Enrichment from Massive Star Clusters
 in II~Zw~40 
}

\author{S. Michelle Consiglio\altaffilmark{1},
  Jean L. Turner\altaffilmark{1},
  Sara Beck\altaffilmark{2},
  David S. Meier\altaffilmark{3}}

\altaffiltext{1} {Department of Physics and Astronomy, University of
  California, Los Angeles, Los Angeles, CA 90095,
  smconsiglio@ucla.edu}

\altaffiltext{2} {Department of Physics and Astronomy, University of
  Tel Aviv, Ramat Aviv, Israel}

\altaffiltext{3} {Department of Physics, New Mexico Institute of
Mining and Technology, Socorro, NM 87801}

\begin{abstract}

We present subarcsecond images of submillimeter CO and continuum emission from a local galaxy
forming massive star clusters: the blue compact dwarf galaxy \iizw. At $\sim$0.4\arcsec\ resolution (20 pc),
the CO(3--2), CO(1--0), 3mm and
870$\mu$m continuum maps illustrate 
star formation on the scales of individual molecular clouds. 
Dust contributes about a third of the 870$\mu$m continuum emission, with free-free accounting for the rest. 
On these scales, there
is not a good correspondence between gas, dust, and free-free emission. Dust continuum is enhanced
toward the star-forming region as compared to the CO emission. We suggest that an unexpectedly
low and spatially variable gas-to-dust ratio is the result
 of rapid and localized dust enrichment of clouds by the massive clusters of the starburst.

\end{abstract}

\section{Introduction}
\label{sec:intro}

Cycling of gas in galaxies drives galactic evolution; massive  stars are
  important in this process because of their high luminosities, intense and
  hard radiation fields, and mass loss.
  In particular, concentrations of young, massive stars in 
clusters can
rapidly create and expel
large masses of metal-rich gas, 
which can potentially produce dust-enriched clouds \citep{2015Natur.519..331T}. 
Dust-enriched gas can affect gas dynamics and 
  interactions of young stars with their surroundings, 
  determining  the manner in which gas and dust
  are dispersed from massive clusters
  \citep{2004ApJ...610..226S, 2006ApJ...643..186T,2009ApJ...703.1352K,2010ApJ...709..191M,2011ApJ...732..100D, 2016ApJ...819..137K}.
  Differences between 
  CO-derived and dust-derived masses in clouds 
of the LMC \citep{2013ApJ...774...73I, 2016MNRAS.456.1767G} suggest local 
  variations in either the CO-to-\htwo\ conversion factor or in the gas-to-dust ratios (GTDs). 

Previous submillimeter  
 observations of  modest spatial resolution blend 
 regions of different ages  and enrichment,  
which might explain globally-uniform GTDs that are observed   
  \citep{2015ApJ...799...96G}.
  The Atacama Large
  Millimeter/Submillimeter Array (ALMA) has made possible  high resolution submillimeter maps which
  can resolve  individual
    giant molecular clouds in local galaxies.
  It is now  possible to test the assumption that gas and dust are well mixed on cluster scales.

\iizw\ is a nearby (D=10~Mpc), low-metallicity \citep[Z=0.25~Z$_\odot$,][]{2009ApJ...696.1668B}, 
blue compact dwarf galaxy. 
 Although it hosts an exceedingly 
 luminous  ``extragalactic HII
 region"  \citep{1970ApJ...162L.155S, 1981ApJ...247..823T, 1983ApJ...273...81K},
 it has very faint CO lines 
\citep{1992A&A...265...19S, 2001AJ....121..740M, 2005A&A...438..855I}.  
Its region of intense star formation
 \citep{BT02, 2006ApJ...639..157W,
 2008A&A...486..393V,2013ApJ...767...53B, 2014AJ....147...43K, 2016arXiv160701779K}
 contains at least two massive star clusters of $\sim 10^5$--$10^6~M_\odot$ each
 \citep{2013ApJ...767...53B}. 
This ``starburst" dominates the infrared spectrum of \iizw,  which has 
an unusually  warm peak at 60$\mu$m \citep{1993AJ....106.1743V}. 
The dominance of the starburst  makes 
 II Zw 40  
an excellent target for studying the effects of massive clusters on their environments.

We present here some of the first ALMA submillimeter $\lambda$3mm and $\lambda$870$\mu$m 
continuum and CO observations of an extragalactic star-forming region with
massive young star clusters.  We have mapped CO(3-2), CO(1-0), and continuum in \iizw.
From these images we can study 
how well correlated 
  CO and submillimeter
  dust continuum emission are on the scales of star-forming regions. 
 
\section{Observations}
\iizw\ (UGCA116) was observed in ALMA Bands 3 and 7 
as a Cycle 2 (Early Science)
program (ID = 2013.1.00122.S, PI = J. Turner). Band 7 observations took place on 2014 13 August and 13 December. 
 A single field with an 18\arcsec\ field-of-view centered on
05:55:42.620, +03.23.32.0 
was observed with 2844 seconds on source. The uv range covers $\sim$17-983 k$\lambda$; 
structures $\gtrsim$12\arcsec\
are undersampled. This is not a problem for CO(3-2), since
emission in individual channels is compact, nor for free-free emission.
However, it could be
an issue for dust, which traces both atomic and molecular gas. 
Band 3 observations took place on 2015 September 2 and 3, with 
4927 seconds on source.
Spectral resolution 244.141
kHz per channel resolves both CO(3--2) and CO(1--0) lines; we present 
the integrated intensity maps for emission $>$1.5$\sigma$. The uv range for the 2.6mm  
images covers $\sim$5-525k$\lambda$, sampling structures up to $\sim$40\arcsec ($\lesssim$1.9~kpc).
Bandpass, phase, and flux were calibrated
with J0607-0834, J0532+0732, and J0510+180 respectively.
Data calibration using CASA  followed the pipeline.
Continuum emission was subtracted in the (u,v) plane before making line maps.
All maps are convolved to a 0.47\arcsec\ x 0.4\arcsec, PA=51.02\degree\ beam.

For the continuum maps, RMS noise is  $35\mu$ \jpb\ for the 2 GHz band centered at 345.796 GHz (``870$\mu$m"), and
20 $\mu$\jpb\ at 115.27 GHz (``3 mm").
The RMS in a single 2 \kms\ Hanning-weighted channel  is 1.5 m\jpb\ for  
CO(1-0) maps and 1.6 m\jpb\ for 
CO(3-2). The RMS in the integrated intensity maps is
19 m\jpbkms\ in CO(1--0) with a 1.5$\sigma$ cutoff, and 14 m\jpbkms\ for CO(3--2) with a 2.5$\sigma$ cutoff.
 Single dish fluxes from the 
literature are 1.4-4.6 Jy \kms\ for CO(1-0) in a 22-24\arcsec\ beam \citep{1992A&A...265...19S, 2004A&A...414..141A}
and $<$6.7 Jy \kms\ for CO(3-2) \citep{2001AJ....121..740M,2005A&A...438..855I}. 
We may detect up to half of the CO(1-0) single dish flux, but this is unclear from the conflicting published values.
We detect most, perhaps all of the CO(3--2) single dish flux, 
based on comparison with single dish upper limits. 

\section{3mm free-free continuum, HII regions and star formation rate in \iizw}

Free-free emission from the HII region dominates the 3mm continuum of
\iizw, shown in Figure~1b; synchrotron and dust contributions are negligible. 
The 3mm  flux density for the inner 2-3\arcsec\
 is $S_{3mm}=7.6\pm$0.2 mJy, with a peak of
1.67$\pm 0.02$ m\jpb\  at RA: 05:55:42.614 Dec: 03.23.32.01 (J2000).  
Using opacity and recombination coefficient expressions 
from Draine (2011a,b) \nocite{2011piim.book.....D,2011ApJ...732..100D} we find,  
$N_{Lyc} = 1.25 \times 10^{50}~T_4^{-0.51}~\nu_{11}^{+0.118}~\Big(\frac{n_i}{n_p}\Big) 
~\rm D_{Mpc}^2~S(mJy)$
and for T$_{e}$= 13000~K \citep{WR93,2005ApJS..161..240T}, 
we obtain a Lyman continuum rate of $N_{Lyc}=8.3\pm0.2\times 10^{52}\,\rm s^{-1}$. 
This is higher than estimated from cm-wave fluxes \citep{2014AJ....147...43K}, 
which can be caused by  particularly 
 dense ionized gas turning optically thin at millimeter 
 wavelengths \citep{BT02}. 
Since dust competes with gas for UV photons the 
 actual Lyman continuum rate will be higher than this value. 
Following 
\citet{2001ApJ...555..613I}, based on the observed $L_{IR}\sim 1.9\times 10^9~\rm L_\odot$ 
\citep[][corrected for distance]{2008ApJ...678..804E}
and  $N_{Lyc} $ we estimate that 72\%\ 
of the Lyman continuum photons end up ionizing hydrogen.
The corrected Lyman continuum rate is therefore,
 $N_{Lyc}^{corr}=1.2\pm0.2\times 10^{53}\,\rm s^{-1},$ corresponding
to $\sim$12000 O stars, and total stellar mass $\sim 1-2\times10^6$~\msun.

\section{Dust and Molecular Gas in \iizw}
\subsection{870$\mu$m continuum and dust distribution in \iizw}
\begin{figure}
  \includegraphics[width=\linewidth]{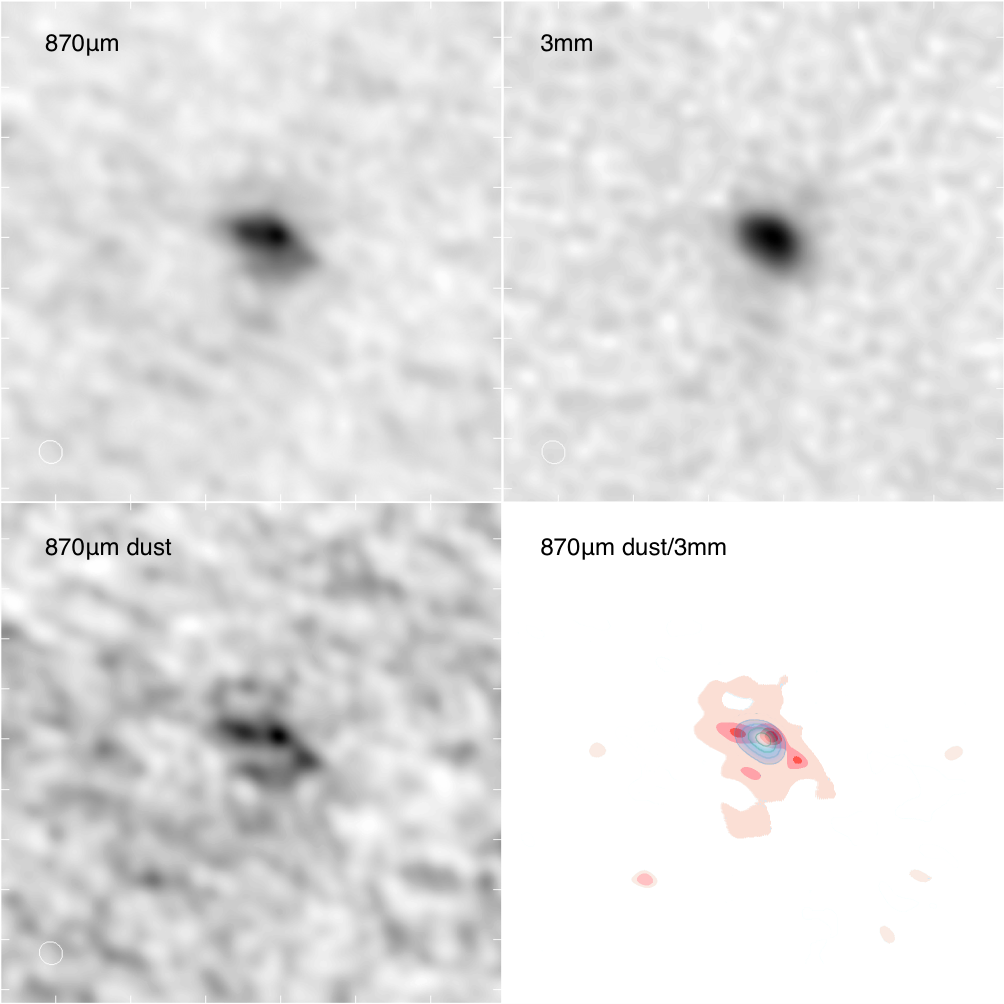}
  \caption{{ALMA continuum and dust maps of \iizw. Beams are shown at
      lower left
      and are all 0.47\arcsec x0.44\arcsec, PA 51.0\degr.
      Tick marks, in white, are 0.5\arcsec. 
      \it(top left) } 870$\mu$m map. The RMS is $\sim$0.035~m\jpb, peak flux is 1.74~m\jpb\ at
    05:55:42.612, 03:23:32.02 (J2000). Total flux 
    is 9.7$\pm$0.5~mJy.
 { \it(top right)} 3mm map.  The RMS is $\sim$0.02~m\jpb, 
  peak flux is 1.67~m\jpb\ at
  05:55:42.614, 03:23:32.01 (J2000).   Total flux is 7.6$\pm$0.2 Jy. 
 {\it(bottom left)} 870$\mu$m dust map (see text.) The RMS is $\sim$0.035~m\jpb, peak flux
  is 0.38~m\jpb. Total dust flux is 3.0$\pm$0.5~mJy. 
  {\it(bottom right)} Schematic overlay of the 870$\mu$m dust in orange with 
  3mm free-free emission (HII region) in blue. Greyscale stretch is
  linear: from -1.5$\sigma$ (white) to peak intensity (black).
  }
  \label{fig:all_maps}
\end{figure}

The 870$\mu$m continuum image is shown in Figure~\ref{fig:all_maps}a; 
 total flux for the inner 6\arcsec\ is
$S_{870\mu m}= 9.7\pm0.5$~mJy and the peak is 1.74$\pm$0.06 m\jpb.  
\citet{2013MNRAS.429.3390H}
found $13.6\pm 2.0$~mJy, at 880$\mu$m in a map with the Submillimeter Array. 
 His peak
flux of 8.5 m\jpb\ in a 7\arcsec\ beam is consistent with our total flux measurement for a
region of the same size. 
 This is $\sim$10\% of the flux in the single dish JCMT-SCUBA maps, much of which
 arises in the
   extended  HI tail 
\citep{2005A&A...434..867G,2005A&A...434..849H,VZ98}.  
The 870$\mu$m continuum includes
both dust and free-free emission.   
To map only the dust, we extrapolate the 3mm free-free emission
 to 870$\mu$m using the spectral index -0.118 \citep{2011piim.book.....D}.  
The total free-free flux density of $S_{870\mu m}^{ff}=6.7\pm0.2$~mJy is $\sim$two-thirds of the total.
The remaining flux density is due to dust; we obtain $S_{870\mu m}^{dust}= 3.0\pm0.5$~mJy. 
This value agrees with that of \citet{2016arXiv160701779K}. The contribution of dust to
the 3mm continuum, for dust emissivity $\beta = 1.7$, is $S_{3mm}^{dust}<0.1$ mJy.

 Subtracting the scaled 3mm free-free continuum from the 870$\mu$m gives a map of the dust emission alone, 
 shown in Figure 1c.
Dust is concentrated near the star formation in a region
$\sim$2\arcsec, or 100 pc in extent. Dust and ionized gas are not perfectly co-extensive: 
the peak flux in the 870$\mu$m continuum is at 
05:55:42.612, 03:23:32.02 (J2000),  slightly (0.14\arcsec, or 7 pc) shifted from the 3mm peak.
Dust peaks include: a central peak, 0.38 m\jpb\ (42s.604, 32\arcsec.06),  western peak,
0.28 m\jpb\ (42s.564, 31\arcsec.55),  eastern peak, 0.32 m\jpb\ (42s.649,
32\arcsec.11), and  southern peak, 0.24 m\jpb\ (42s.632, 31\arcsec.33). 
Dust fluxes in Table 1 differ  from these because those are measured at  
 CO(3-2) peaks.

\subsection{Molecular Gas Distribution and Conditions}

Integrated line images of CO(3-2) and
CO(1-0) are shown in Figure 2.
\begin{figure}
  \includegraphics[width=\linewidth]{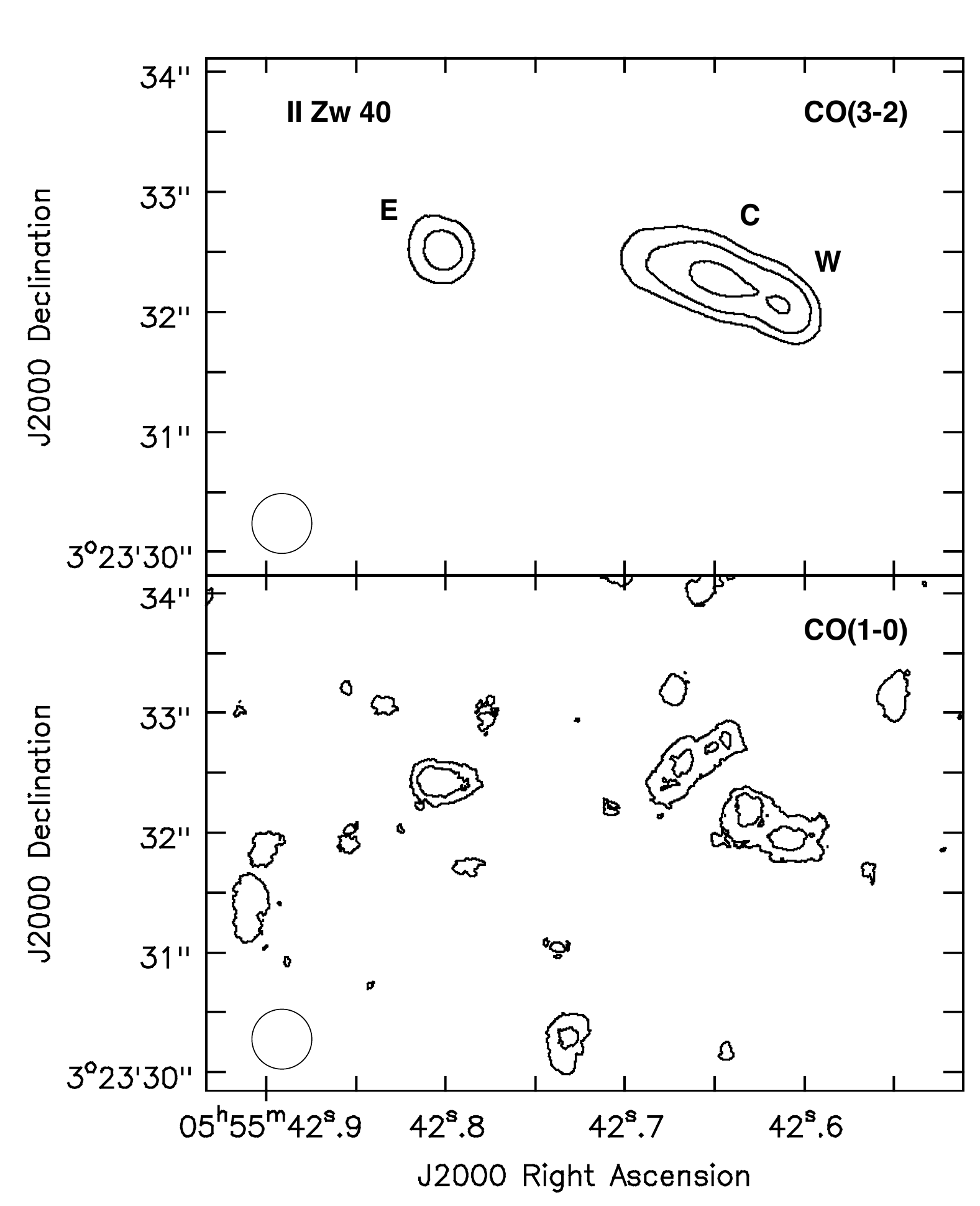}
  \caption{{\it (top)} CO(3-2) and {\it (bottom)} CO(1-0) integrated line intensity in \iizw. 
  Contour levels
    are 0.37, 0.52, 0.74~\jpbkms\ ($\sim$26, 37, 53$\sigma$) for CO(3-2)
    and 0.05, 0.071, 0.10~\jpbkms\ ($\sim$2.6, 3.7, 5$\sigma$) for CO(1-0). 
    Contour levels
  are 7.4 times higher for the CO(3-2) plot, as 
   CO(3-2) is $\sim$7.4 times stronger than
  CO(1-0) for optically
  thick, thermal emission at 30K. 
  }
  \label{fig:CO}
\end{figure}
CO(3-2) traces dense ($\sim$10$^{4}$ cm$^{-3}$) gas, so the clouds in these images
are dense clouds associated with the star formation. 
We restrict our discussion to the CO(3--2) and CO(1--0) in the dense, compact
CO(3-2)-emitting clouds in this paper; 
there may be more extended and
diffuse clouds traced by CO(1-0) that may not be well-represented in these images, while we recover
most of the CO(3-2) flux ($\S$ 2). 
CO(3-2) has been mapped previously by \citet{2016arXiv160701779K}, but 
  they did not detect CO(1-0) with $\sim$3 times higher rms.

   The  overall integrated flux ratio of CO(3--2) and CO(1--0), $R_{31}$, 
   where $I_{line}=
\int S_{CO} dv$ (Jy \kms), is on average
$R_{31}=I_{32}/I_{10}=(4.1\pm1.0/0.5\pm0.2) \sim 8$, 
and ranges from \rthreeone$\sim$3--13  
for CO(1--0) integrated intensity $>$1.5$\sigma$.
 The line ratio in thermal, optically thick gas is 
\rthreeone $=$7.4 for $T_{gas}=30$~K. 
The observed \rthreeone\ is consistent with thermalized, warm gas. 
We have chosen the contours for the CO(3-2) in Figure 2 to be 7.4 times those of CO(1-0) to
demonstrate the similarities in the maps at this value of \rthreeone. 
The central extended cloud has significantly elevated \rthreeone $\gtrsim 20$, 
as does the western cloud near the 3mm continuum
 peak (\rthreeone$\sim$10), indicative of warmer gas directly associated with star formation.
 The observed \iizw\ \rthreeone\ cloud values are much higher than the galactic $<$\rthreeone$>$=1.6
 found in the JCMT survey of local SINGS galaxies \citep{2012MNRAS.424.3050W}.
 Since we identify clouds
 based on CO(3-2) structure, it is not surprising that these clouds have higher
 than average \rthreeone\ values. 
These CO(3-2)-emitting clouds are denser and warmer in \iizw\ than typical 
spiral disk clouds,
with $\log (n/\rm cm^{-3})\gtrsim$3.5-4. 

\subsection{Gas-to-Dust in \iizw}
Figure 3 shows the gas and dust together.
\begin{figure}
  \includegraphics[width=\linewidth]{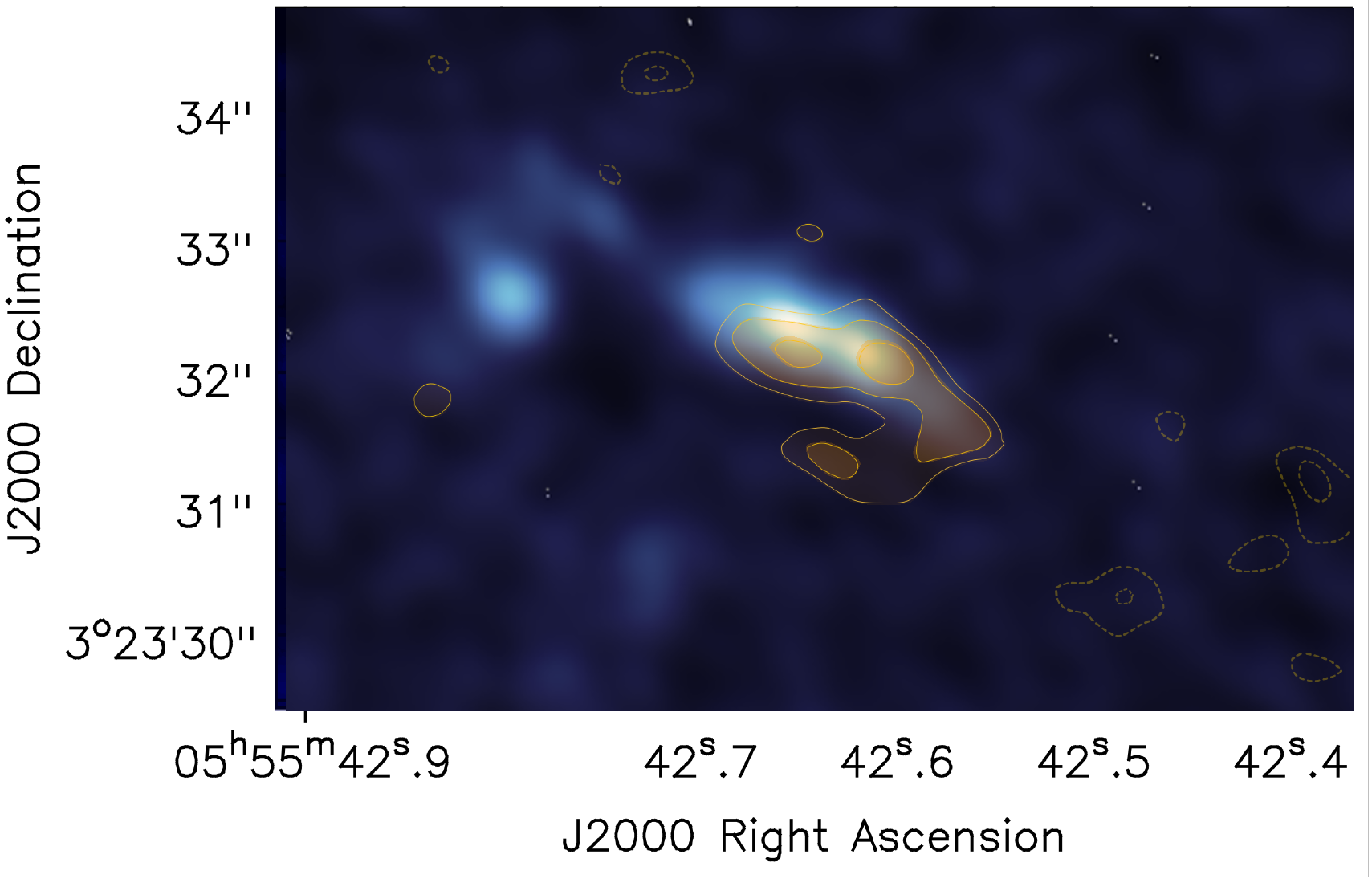}
  \caption{CO(3-2) emission (blue) with the dust map (orange) to show the relative spatial
  relation of the CO and dust emission. 
  CO(3-2) map monochrome scale is from 3$\sigma$ = 0.05
  to its peak, 0.85~(\jpbkms). The dust continuum contours are $\pm$0.15, 0.21, 0.3~m\jpb, with
  negative contours dashed.
  }
    \label{fig:CO_dust}
\end{figure}
The CO-to-dust emission ratio varies significantly 
across the starburst region.  
The isolated CO cloud to the east is molecular gas rich, with almost no
dust emission. In contrast, the westernmost dust peak 
is relatively 
dust-rich, with weak CO.  The ratio of CO(3-2) line flux to dust continuum flux varies from 
$S_{CO32}/S_{870}^{dust}\sim 1400$~\kms\  to $S_{CO32}/S_{870}^{dust}\sim 13000$~\kms, almost an order
of magnitude.  

We calculate \gtd\ by determining
dust and gas masses separately. We focus on three regions 0.6\arcsec\ in
diameter: ``Cloud W", coincident with the 3mm free-free peak, ``Cloud C", to the
immediate east of the free-free peak, and ``Cloud E", an isolated
cloud to the east (Figure 2). Clouds E and
W are similar in CO flux, but very different in dust emission.
 To obtain dust mass, we use a dust opacity calculated for an extended region in the LMC by \citet{2011A&A...536A..88G},
  with $\kappa_{870\mu m}=0.9~\rm
cm^2\,g^{-1}$ and $\beta=1.7$. 
For Clouds C and W, we use the 33K IRAS 60/100
color temperature \citep{2008ApJ...678..804E} and for the isolated 
Cloud E, we use T$_{Dust}$=20K. 
We obtain a total dust mass for the central 5\arcsec\ 
of 17000$\pm$5000~\msun (for T=33K). This may underestimate the true dust mass, since
the SED peak is dominated by the warmest dust, which will not dominate the mass.

Gas masses are derived from CO(1-0) using 
the conversion factor, $X_{CO}$. Since the CO(1-0) detection is weak,
we also include
 masses based on the CO(3-2)
 fluxes for comparison, using \rthreeone=7.4 to obtain the equivalent CO(1-0)
  flux. Because of excitation effects in CO(3-2), the CO(1-0) masses, while uncertain, 
  are more reliable for using $X_{CO}$.
   We adopt the LMC conversion factor of
$X_{CO}=4.7\times10^{20}~\rm cm^{-2}\,(\rm K ~km\, s^{-1})$ \citep[]{2003A&A...401...99I,2010MNRAS.406.2065H}. 
This value was obtained for more diffuse gas which is more likely to have CO-dark \htwo; it 
could well overestimate the molecular gas in these dense clouds. 
Gas and dust masses and \gtd\  calculated for the three Clouds E, C, and W, associated with CO(3-2) peaks, and
for the entire central 5\arcsec\ starburst region are given in Table 1. The overall gas mass based on
CO(1-0) is $M_{gas}=1.1\times10^6$~\msun, including helium.

\gtd\ is expected to scale with galactic metallicity, 
although with large scatter \citep{2014A&A...563A..31R}.   
The metallicity of \iizw\ is 
12+log(O/H)$\sim$8.1, $\sim$0.25 Z$_{\odot}$ \citep{2009ApJ...696.1668B}, 
 so the predicted \gtd\  is 
  $\sim$400-600.  Only Cloud E approaches
this value. The observed  values of \gtd $\sim$60-170 
are far lower than expected for this low metallicity galaxy.

\begin{deluxetable*}{lccccccccccc}
	\tabletypesize{\scriptsize}
	\tablecaption{Gas, Dust and GTD}
	\tablewidth{0pt}
	\tablenum{1}
	\tablehead{\colhead{Name}&%
          \colhead{Position}&%
          \colhead{CO(1-0)}&%
          \colhead{CO(3-2)}&%
          \colhead{$\frac{(3-2)}{(1-0)}$}&%
          \colhead{$M^{(1-0)a}_{gas}$} &%
          \colhead{$M^{(3-2)a}_{gas}$} &%
          \colhead{Dust Flux$^{b}$} &%
          \colhead{$M_{dust}^{b}$}&%
          \colhead{GTD$^{c}$}\\
	  \colhead{}&%
          \colhead{05:55; +03:23}&%
          \colhead{(Jy~km~s$^{-1}$)}&%
          \colhead{(Jy~km~s$^{-1}$)}&ratio&%
          \colhead{($10^{4}M_{\odot}$)}&%
          \colhead{($10^{4}M_{\odot}$)} &%
          \colhead{(mJy)}&%
          \colhead{($M_\odot$)}&%
          \colhead{(1-0), (3-2)}}

\startdata
Cloud  W$^d$ & 42.609, 32.03 & 0.06 & 0.71 & 12 & 13 & 21  & 0.34 & 1940 & 69, 110\\
Cloud E$^d$ & 42.802, 32.482 & 0.055 & 0.52 & 9 & 12 & 16  & $<$0.04 & $<$450 & $>$270, $>$350\\
Cloud C$^d$ & 42.647, 32.264 & 0.045 & 0.93 & 21 & 10 & 28 & 0.29 & 1650 & 60, 170\\
Starburst$^e$ & 42.695, 31.763 & 0.50 & 4.1  & 8 & 110 & 120  & 3.0 &  $1.7\times10^4$ & 65, 72\\

\enddata
\tablenotetext{a}{From observed CO(1--0) flux and CO(3--2) flux with \rthreeone=7.4. Integrated
  line map RMS is 19 and 14 m\jpbkms\ for CO(1--0) and CO(3--2) respectively. Mass includes He.}
\tablenotetext{b}{Dust fluxes are measured from regions defined by the
  CO, not the dust peaks themselves, this underestimates the dust. Masses are estimated using a dust
  temperature of 33K except for Cloud E, which is 20K. Cloud E is an upper limit of 3$\sigma$.}
\tablenotetext{c}{GTDs computed from CO(1-0) masses on
  the left and CO(3-2) masses on the right. }
\tablenotetext{d}{Fluxes for 0.6\arcsec~diameter regions}
\tablenotetext{e}{For a 5\arcsec~diameter region }

\end{deluxetable*}

\section{Low and Spatially Varying Gas-to-Dust:  Gas Conditions or Starburst Dust Production?}
 Derived \gtd\ values in \iizw\  are consistently   lower than
 predicted from its metallicity. 
 \gtd\ also varies significantly across the region; from 70
 in Cloud W to $>$270 in Cloud E, 
 clouds 
 similar in 
 CO flux 
 and size. 
 There are several possible explanations for the
 variation.
The missing gas mass is unlikely to be HI in these dense clouds. 
It is difficult to compare to the  large beams
of HI maps, but based on observed column densities \citep{VZ98}, we estimate that roughly 10\% of the
mass could be in HI in these compact clouds. Columns  higher than this will become molecular
gas. CO-dark \htwo\ is unlikely to contribute significantly in dense clouds \citep{2014A&A...561A.122L};
 the adopted
conversion factor has already accounted for this to some extent. 
$X_{CO}$ variations could explain the variable \gtd\ \citep[as suggested
for the LMC,][]{2016MNRAS.456.1767G}. For the entire region to have the same \gtd\ as Cloud E 
would require
 $X_{CO}=35$-$45\times10^{20}~\rm cm^{-2}\,(\rm K ~km\, s^{-1})$ in dense
clouds where CO-dark \htwo\ is not expected, and it 
  requires
that  $X_{CO}$ vary by a factors of 4-10 across a region of  $\lesssim$200 pc extent.  
High values of $X_{CO}$ have been inferred in
other low metallicity galaxies, but under the assumption of a uniform \gtd\ \citep{2014Natur.514..335S,2015ApJ...804L..11S}, which
we argue below is questionable. 
We also note the apparent contradiction in any model relying on CO-dark \htwo: to explain
these results 
requires extinction be lowest 
in clouds where dust emission is strongest.  

We favor instead for  \iizw\ a model in which 
the \gtd\ is low and variable because the starburst has enriched the
surrounding clouds in metals and dust. 
Massive clusters such as those in \iizw\
contain the most massive O stars \citep{2016ApJ...823...38S}
that lose most of their mass within a few Myr \citep{2012A&A...537A.146E}.
Both the Geneva high mass loss stellar models and the rotating models 
used in  Starburst99 \citep{1999ApJS..123....3L, 2014ApJS..212...14L} predict metal (CNOSiMg) yields 
of up to 60,000~\msun\ for clusters of this luminosity ($\sim 2\times 10^9~\rm L_\odot$)
that are 3 Myr or older.  If half of
this metal yield were in the form of dust, as in the Galaxy \citep{2009ApJ...700.1299J}
then 30,000~\msun\  of dust could be present. We infer a total dust mass that is
half this amount. In this scenario, the dominant source of dust is production in the stars
of the massive clusters, and the contribution
of dust from the natal clouds is small. This explains the contrast in \gtd\ between Cloud W, which has been 
enriched by the massive clusters, and Cloud E, which has little star formation. 
This theory is also consistent with the significant spatial variations seen in the inferred oxygen abundances
within this region by \citet{2009ApJ...696.1668B}. Enrichment is a natural and expected result of 
the stellar evolution of large clusters containing O stars. 

Localized dust enrichment 
also provides an explanation for the spatial distribution of the dust
\citep{2011ApJ...732..100D}.
 In luminous clusters, radiation pressure can cause the dust to drift relative
 to the gas, by $\sim$1~\kms. 
  This could explain the slight 
offset of the main dust peak ($\sim$10 pc) from the radio continuum peak and presence of dust mostly outside
the HII region. Regions of dust
 emission without gas or continuum, such as the westernmost dust peak, may signal the presence of  
 older clusters which no longer have HII regions. At a sound speed of 10 \kms,
 100 Myr is required to cover the observed dust-emitting region; the presence of radio continuum emission
 suggests that these clusters are much younger. The starburst in \iizw\ is young enough
 that the cluster-enriched gas has not had time to fully disperse.
 
These observations of \iizw\ suggest that dust enrichment from massive clusters can be rapid and 
localized in starburst galaxies. This could 
explain why very high values of $X_{CO}$ are inferred when Galactic \gtd\ are
assumed \citep[e.g.,][]{2014Natur.514..335S,2015ApJ...804L..11S}. 
Localized enrichment by massive star clusters 
may also explain why dusty galaxies, or dusty regions within galaxies,
 are seen so early in the universe \citep{2015Natur.519..327W}.  

\section{Summary and Conclusions}

We present ALMA continuum maps at 3mm and 870$\mu$m and integrated intensity images
of  the CO(3--2) and CO(1--0) lines in the 
blue compact dwarf galaxy \iizw, at a resolution of $\sim$0.4\arcsec\ ($\sim$20 pc).
Based on a 3mm continuum flux of 7.6$\pm$0.2~mJy, we obtain 
a Lyman continuum rate of $N_{Lyc}^{corr}=1.2\pm0.2\times 10^{53}\,\rm s^{-1}$ 
with $\sim$72\% of the uv photons ionizing the gas. From 3mm and 870$\mu$m
continuum maps, we construct a dust-only continuum map. 
For a dust flux of 
$S_{870\mu m}^{dust}=3.0\pm0.5$~mJy, 
 we estimate a dust mass of $\sim17,000\pm 5000$~\msun, for
 T$_{dust}$=33 K. 
  While CO(1-0) emission is weak, CO(3-2) and CO(1-0) line emission
  maps show similar structure and are consistent with warm, thermalized gas: 
 CO(3-2) integrated fluxes are $\sim$3-13 times stronger than the CO(1-0).
 Adopting an LMC $X_{CO}=4.7\times10^{20}~\rm cm^{-2}\,(K\,$\kms$)^{-1}$,
 we estimate a gas mass of $M_{gas}=1.1\times 10^6$~\msun, including helium, comparable
 to the masses of the  star clusters. 
  The CO(3-2) extends over 4\arcsec\ (~200~pc), significantly larger than the free-free emission, 
   while the dust is concentrated to the immediate $\sim$2\arcsec\ (100~pc) vicinity of the starburst.

The ratio of CO(3-2) line flux to dust continuum flux  varies from 
$S_{CO32}/S_{870}^{dust}\sim 1400$~\kms\  to $S_{CO32}/S_{870}^{dust}\sim 13000$~\kms\ across
the source.
Some of this variation is
due to CO(3-2) excitation, but much of it is due to gas column
variations.
 For values of $X_{CO}$ and dust opacity from the LMC, 
 we find the gas-to-dust ratio, based on CO(1-0) fluxes, is unexpectedly low 
 and variable across the 200 pc region:  \gtd$\sim$60-70 in the 0.6\arcsec\  regions ($\sim$50 pc)
 adjacent to the starburst, but \gtd$>$270 in a  similar, but isolated and  
quiescent cloud  $\lesssim$200 pc to the
east. 
We argue that a low and variable \gtd\ ratio 
in \iizw\ is a result of 
dust produced by the O  
stars within the massive star clusters, 
an example of localized dust enrichment by a starburst that has not yet had time to disperse.

\acknowledgments

This paper makes use of the following ALMA data: ADS/JAO.ALMA\#2013.1.00122.S. 
ALMA is a partnership of ESO (representing its member states), NSF (USA) and NINS (Japan), together with NRC (Canada), NSC and ASIAA (Taiwan), and KASI (Republic of Korea), in cooperation with the Republic of Chile. The Joint ALMA Observatory is operated by ESO, AUI/NRAO and NAOJ. The National Radio Astronomy Observatory (NRAO) is a facility of the National Science Foundation operated under cooperative agreement by Associated Universities, Inc.  Support for this work was provided by the NSF through award GSSP SOSPA2-016 from the NRAO to SMC and grant AST 1515570 to JLT, and by the UCLA Academic Senate through a COR seed grant. We
thank the anonymous referee for thorough comments that aided our  manuscript.

{\it Facilities:} \facility{ALMA}.


\begin{thebibliography}{}
\bibitem[Albrecht et al.(2004)]{2004A&A...414..141A}
  Albrecht, M., Chini, R., Kr{\"u}gel, E., M{\"u}ller, S.~A.~H., \& Lemke, R.\ 2004, \aap, 414, 141 
\bibitem[Beck et al.(2013)]{2013ApJ...767...53B} 
Beck, S., Turner, J., Lacy, J., Greathouse, T., \& Lahad, O.\ 2013, \apj, 767, 53 
\bibitem[Beck et al.(2002)]{BT02}
Beck, S., Turner, J., Langland-Shula, L., Meier, D., Crosthwaite, L., \& Gorjian, V. 2002, AJ, 124, 2516 
\bibitem[Bordalo et al.(2009)]{2009ApJ...696.1668B} 
Bordalo, V., Plana, H., \& Telles, E.\ 2009, \apj, 696, 1668 
\bibitem[Draine(2011a)]{2011ApJ...732..100D} 
Draine, B.~T.\ 2011, \apj, 732, 100 
\bibitem[Draine(2011b)]{2011piim.book.....D} 
Draine, B.~T.\ 2011, 
Physics of the Interstellar and Intergalactic Medium by 
Bruce T.~Draine.~Princeton University Press, 2011.~ISBN: 978-0-691-12214-4,  95
\bibitem[Engelbracht et al.(2008)]{2008ApJ...678..804E} 
Engelbracht, C.~W., Rieke, G.~H., Gordon, K.~D., et al.\ 2008, \apj, 678, 804-827 
\bibitem[Ekstr{\"o}m et al.(2012)]{2012A&A...537A.146E} 
Ekstr{\"o}m, S., Georgy, C., Eggenberger, P., et al.\ 2012, \aap, 537, A146 
\bibitem[Galametz et al.(2016)]{2016MNRAS.456.1767G} 
Galametz, M., Hony, S., Albrecht, M., et al.\ 2016, \mnras, 456, 1767 
\bibitem[Galliano et al.(2011)]{2011A&A...536A..88G} 
Galliano, F., Hony, S., Bernard, J.-P., et al.\ 2011, \aap, 536, A88 
\bibitem[Galliano et al.(2005)]{2005A&A...434..867G} 
  Galliano, F., Madden, S.~C., Jones, A.~P., Wilson, C.~D., \& Bernard, J.-P.\ 2005, \aap, 434, 867
\bibitem[Groves et al.(2015)]{2015ApJ...799...96G}
  Groves, B.~A., Schinnerer, E., Leroy, A., et al.\ 2015, \apj, 799, 96
\bibitem[Hirashita(2013)]{2013MNRAS.429.3390H} 
Hirashita, H.\ 2013, MNRAS, 429, 3390
\bibitem[Hughes et al.(2010)]{2010MNRAS.406.2065H}
Hughes, A., Wong, T., Ott, J., et al.\ 2010, \mnras, 406, 2065
\bibitem[Hunt et al.(2005)]{2005A&A...434..849H} 
 Hunt, L., Bianchi, S., \& Maiolino, R.\ 2005, \aap, 434, 849
\bibitem[Indebetouw et al.(2013)]{2013ApJ...774...73I}
  Indebetouw, R., Brogan, C., Chen, C.-H.R. et al. \ 2013, \apj, 774, 73
\bibitem[Inoue et al.(2001)]{2001ApJ...555..613I}
 Inoue, A.~K., Hirashita, H., \& Kamaya, H.\ 2001, \apj, 555, 613 
\bibitem[Israel(2005)]{2005A&A...438..855I} 
Israel, F.~P.\ 2005, A\&A, 438, 855 
\bibitem[Israel et al.(2003)]{2003A&A...401...99I} 
Israel, F.~P., de Graauw, T., Johansson, L.~E.~B., et al.\ 2003, \aap, 401, 99
\bibitem[Jenkins(2009)]{2009ApJ...700.1299J}
  Jenkins, E.~B.\ 2009, \apj, 700, 1299
\bibitem[Kepley et al.(2014)]{2014AJ....147...43K} 
Kepley, A.~A., Reines, A.~E., Johnson, K.~E., \& Walker, L.~M.\ 2014, \aj, 147, 43 
\bibitem[Kepley et al.(2016)]{2016arXiv160701779K} 
Kepley, A.~A., Leroy, A.~K., Johnson, K.~E., Sandstrom, K., \& Chen, C.-H.~R.\ 2016, arXiv:1607.01779 
\bibitem[Kim et al.(2016)]{2016ApJ...819..137K} 
Kim, J.-G., Kim, W.-T., \& Ostriker, E.~C.\ 2016, \apj, 819, 137 
\bibitem[Krumholz \& Matzner(2009)]{2009ApJ...703.1352K} 
Krumholz, M.~R., \& Matzner, C.~D.\ 2009, \apj, 703, 1352 
\bibitem[Kunth \& Sargent(1983)]{1983ApJ...273...81K}
Kunth, D. \& Sargent, W.L.W. \ 1983, \apj, 273, 81
\bibitem[Langer et al.(2014)]{2014A&A...561A.122L} 
Langer, W.~D., Velusamy, T., Pineda, J.~L., Willacy, K., \& Goldsmith, P.~F.\ 2014, \aap, 561, A122 
\bibitem[Leitherer et al.(2014)]{2014ApJS..212...14L} 
Leitherer, C., Ekstr{\"o}m, S., Meynet, G., et al.\ 2014, \apjs, 212, 14 
\bibitem[Leitherer et al.(1999)]{1999ApJS..123....3L} 
Leitherer, C., Schaerer, D., Goldader, J.~D., et al.\ 1999, \apjs, 123, 3 
\bibitem[Meier et al.(2001)]{2001AJ....121..740M} 
Meier, D.~S., Turner, J.~L., Crosthwaite, L.~P., \& Beck, S.~C.\ 2001, AJ, 121, 740 
\bibitem[Murray et al.(2010)]{2010ApJ...709..191M} 
Murray, N., Quataert, E., \& Thompson, T.~A.\ 2010, \apj, 709, 191 
\bibitem[R{\'e}my-Ruyer et al.(2014)]{2014A&A...563A..31R} 
R{\'e}my-Ruyer, A., Madden, S.~C., Galliano, F., et al.\ 2014, \aap, 563, A31
\bibitem[Sage et al.(1992)]{1992A&A...265...19S} 
Sage, L.~J., Salzer, J.~J., Loose, H.-H., \& Henkel, C.\ 1992, \aap, 265, 19 
\bibitem[Sargent \& Searle(1970)]{1970ApJ...162L.155S}
Sargent, W.L.W. \& Searle, L. \ 1970, \apj, 162, 155
\bibitem[Shi et al.(2014)]{2014Natur.514..335S} 
Shi, Y., Armus, L., Helou, G., et al.\ 2014, \nat, 514, 335 
\bibitem[Shi et al.(2015)]{2015ApJ...804L..11S} 
Shi, Y., Wang, J., Zhang, Z.-Y., et al.\ 2015, \apjl, 804, L11 
\bibitem[Silich et al.(2004)]{2004ApJ...610..226S}
Silich, S., Tenorio-Tagle, G., \& Rodr{\'{\i}}guez-Gonz{\'a}lez, A.\ 2004, \apj, 610, 226 
\bibitem[Smith et al.(2016)]{2016ApJ...823...38S} 
Smith, L.~J., Crowther, P.~A., Calzetti, D., \& Sidoli, F.\ 2016, \apj, 823, 38 
\bibitem[Tenorio-Tagle et al.(2006)]{2006ApJ...643..186T} 
Tenorio-Tagle, G., Mu{\~n}oz-Tu{\~n}{\'o}n, C., P{\'e}rez, E., Silich,
S., \& Telles, E.\ 2006, \apj, 643, 186
\bibitem[Thuan \& Izotov(2005)]{2005ApJS..161..240T} 
Thuan, T.~X., \& Izotov, Y.~I.\ 2005, \apjs, 161, 240 
\bibitem[Thuan \& Martin(1981)]{1981ApJ...247..823T}
Thuan, T.X. \& Martin, G.E. \ 1981, \apj, 247, 823
\bibitem[Turner et al.(2015)]{2015Natur.519..331T}
Turner, J.L., Beck, S.C., Benford, D.J., et al. \ 2015, Nature, 519, 331
\bibitem[Vader et al.(1993)]{1993AJ....106.1743V} 
Vader, J.~P., Frogel, J.~A., Terndrup, D.~M., \& Heisler, C.~A.\ 1993, \aj, 106, 1743 
\bibitem[van Zee, Skillman \& Salzer(1998)]{VZ98}
van Zee, L., Skillman, E., \& Salzer, J. 1998, AJ, 116, 1186
\bibitem[Vanzi et al.(2008)]{2008A&A...486..393V} 
Vanzi, L., Cresci, G., Telles, E., \& Melnick, J.\ 2008, A\&A, 486, 393 
\bibitem[Walsh \& Roy(1993)]{WR93}
Walsh, J.R., \& Roy, J.R. 1993, MNRAS, 262, 27 
\bibitem[Watson et al.(2015)]{2015Natur.519..327W} 
Watson, D., Christensen, L., Knudsen, K.~K., et al.\ 2015, \nat, 519, 327 
\bibitem[Wilson et al.(2012)]{2012MNRAS.424.3050W} 
Wilson, C.~D., Warren, B.~E., Israel, F.~P., et al.\ 2012, \mnras, 424, 3050 
\bibitem[Wu et al.(2006)]{2006ApJ...639..157W} 
Wu, Y., Charmandaris, V., Hao, L., et al.\ 2006, \apj, 639, 157 




\end{thebibliography}
\end{document}